\documentclass[]{spie}

\usepackage{amsmath,amsfonts,amssymb}
\usepackage{graphicx}
\usepackage[colorlinks=true, allcolors=blue]{hyperref}
\usepackage{authblk} 

\title{A Path to Resource Optimization and Technological Innovation: Advancing Space and Climate Research with Bidirectional Technologies}

\author[a]{Yixuan Cheng}
\author[a]{Maheen H. Mufti}
\author[a]{Carrie He}
\author[a]{Ying Cong Zuo}
\affil[a]{University of Toronto Aerospace Team, Division of Aerospace Policy, 55 St.George St, Toronto, Ontario, Canada M5S 0C9}

\authorinfo{Correspondence Email: chrisyx.cheng@mail.utoronto.ca, policy.teams@utat.ca}

\begin{document} 

\maketitle

\begin{abstract}
This paper introduces Bidirectional Technologies (BiTs), which is defined as technology that addresses the challenges within the aerospace and the climate sectors simultaneously. BiTs presents a means to meet global development agendas, in particular, the United Nations \textit{2030 Agenda for Sustainable Development} and the \textit{Space2030 Agenda}. These frameworks position aerospace innovations as tools to address climate change, explicitly highlighting the shared challenges between both fields. This overlap presents an underexplored opportunity to develop BiTs. To explore this potential, the study conducts an extensive literature review to examine the challenges within four categories in both Earth and space contexts: life support, energy systems, monitoring and exploratory systems, and novel technologies. Key traits that influence successful development of BiTs are then extracted. Based on these traits, a set of actionable recommendations is proposed to serve as a starting point for policymakers and engineers to adapt technologies beyond single-domain applications. Ultimately, this study presents BiTs as a solution to the interconnected challenges on Earth and in space, offering strategies that advance innovation while contributing to global sustainability efforts.
\end{abstract}

\vspace{10pt}

\noindent
\begin{minipage}[t]{0.55\textwidth}
\textbf{Acronyms/Abbreviations} \vspace{2mm} \\
\begin{tabular}{@{}ll@{}} 
    \textbf{BiTs} & Bidirectional Technologies \\
    \textbf{R\&D} & Research \& Development \\
    \textbf{SDGs} & Sustainable Development Goals \\
    \textbf{STI}  & Science, Technology and Innovation \\
    \textbf{UN}   & United Nations \\
\end{tabular}
\end{minipage}
\begin{minipage}[t]{0.4\textwidth}
\textbf{Keywords} \vspace{2mm} \\
\begin{tabular}{@{}l@{}} 
    \quad Sustainability \\
    \quad Space Policy \\
    \quad Climate Change Policy \\
    \quad Knowledge Transfer  \\
    \quad Resource Optimization \\
\end{tabular}
\end{minipage}

\section{Introduction}
The \textit{2030 Agenda for Sustainable Development}, is composed of 17 Sustainable Development Goals (SDGs) and is designed to place the world on a path toward sustainability \cite{UNspace2030_2024}, while acknowledging science, technology, and innovation (STI) as a “key role in achieving the SDGs” \cite{UN_STIs_2022}. Building on this mandate, international space institutions like the Space2030 Agenda have increasingly sought to align with the international effort by framing space as “a driver of sustainable development” \cite{UNspace2030_2024}. Progress toward these goals are reflected by investments as the global space expenditures reached a record USD135 billion in 2024 \cite{croison_government_2025}, while climate finance to developing nations totaled an unprecedented USD115.9 billion in 2022 \cite{oecd_climate_2024}. 

Despite these commitments, progress has been slow \cite{mishara_world_2024}. This dissonance between substantial investment and limited outcomes signals the need for new approaches in technological development. As the 2030 deadline approaches, it is worth asking whether technologies can be reimagined and designed to confront challenges across both the climate and space frontiers simultaneously.

Notably, technological innovations developed for the challenges of Earth-based climate echo the same constraints faced in space operability such as harsh conditions, scarcity of resources, and performance reliably in high-stake situations. This symmetry highlights an underexplored opportunity for bidirectional technologies (BiTs), defined here as innovations capable of serving both space and climate applications. By addressing shared problems with common solutions, BiTs can reduce duplication, pool resources, and foster exchange of knowledge between fields. Such convergence may also enable novel approaches that neither domain could achieve alone. 

However, this emerging field faces significant limitations. Most agendas and initiatives targeting adaptable technologies focus on one-way transfers from space to Earth, such as the Canadian Space Mining Cooperation developing products from space technologies that benefit Earth \cite{csmc_projects_nodate}.

To address these issues, this paper fills these gaps by proposing a framework for the advancement of bidirectional innovations. First, four shared technological categories across Earth and space and the challenges they face are identified. Second, eight traits that directly respond to these challenges are highlighted. Finally, these traits are turned into actionable steps, thus, bridging the gap between aspirations and real-world practices. The goal of this paper is to lower barriers to entry by providing accessible starting points for diverse stakeholders: for engineers, the framework offers design principles; for policymakers, evaluative tools to guide funding and regulation; and for managers and innovators, logistical and institutional pathways to enable implementation. Based on available knowledge, this paper represents the first study of the requirements of developing and regulating technology shared between space and Earth, marking the first step toward facilitating systematic technological transfer in the space-climate context.

\section{Technology Categories, Shared Challenges, and the Eight Traits of Bidirectional Success}
To better understand the key drivers of cross-domain technology use, this paper examines four categories relevant to both climate and space: life support, monitoring and exploration systems, energy systems, and novelty technologies. For each category, it identifies the main challenges that hinder development. While multiple and often overlapping challenges exist, the discussion focuses on two representative ones in detail for each category.

\subsection{Categories}
\subsubsection{Life Support Technologies}
Life support technologies sustain human life by providing essentials such as food, clean water, and breathable air. These systems are vital not only for maintaining habitability on Earth but also for enabling long-duration space missions. As climate change intensifies environmental disruptions, there is growing demand for technologies that enhance adaptation and resilience. Solutions that emphasize circularity, self-sufficiency, and resource efficiency are especially valuable in vulnerable settings such as arid regions, disaster zones, and remote communities. This convergence in priorities creates opportunities for bidirectional innovation: life support systems developed for space can be adapted to address terrestrial climate challenges, and vice versa. Both domains are shaped by similar constraints—resource scarcity, system reliability, and closed-loop efficiency—while cost-effectiveness remains essential for broader adoption and long-term impact.

\paragraph{Circularity}
Earth-based water filtration systems follow well-established standards for cleanliness and effectiveness, while space-developed systems emphasize closed-loop operation and reliability under extreme conditions, which are traits that align well with climate adaptation needs. In space, water recovery systems like NASA’s Environmental Control and Life Support System (ECLSS) demonstrated the achievement of 98\% of wastewater recovery in 2023, which is key to reducing payload weight \cite{Gaskill_2025}. However, challenges remain in achieving true circularity in Earth-based systems, with existing systems discarding brine and other inputs to water treatment, alongside other political, institutional, and technological challenges in developing and deploying truly “circular” solutions to address freshwater needs on Earth \cite{koseoglu-imer_current_2023}. Potential for exploring how success achieved in space-based water filtration systems can be deployed to achieve similar results in Earth-based applications.

\paragraph{Cost Effectiveness}
Climate-controlled food systems, or Controlled Environment Agriculture (CEA), can support self-sufficient food production in extreme or resource-constrained environments for both space and Earth. However, the major drawbacks for CEA systems come from high energy demands, leading to significant operational costs, and high initial construction fees \cite{10.1093/pnasnexus/pgaf078}. Without concrete solutions to cost effectiveness, it is unlikely to see wide adoption of CEA systems face significant barriers to scaling up and becoming widespread.

\subsubsection{Energy Systems}
Energy systems, comprising all components related to the production, conversion, delivery, and use of energy \cite{pachauri_climate_2015}, play a vital role in terrestrial and aerospace applications. On Earth, the growing global awareness of climate change has driven the development and adoption of renewable energy sources, such as solar, wind, and biomass energy, offering numerous environmental and economic benefits \cite{ang_comprehensive_2022}. However, its development and implementation face significant technological and adaptational challenges. 

As a relatively new sector, renewable energy suffers from a lack of established infrastructure, sufficient investment, and public trust compared to the more mature, fossil fuel-based systems, with energy reliability as another major concern \cite{ang_comprehensive_2022}, \cite{bhuiyan_overcome_2022}. To address this limitation, grid-scale energy storage (GSES) systems are essential for ensuring stable power delivery while achieving reduced emissions \cite{ang_comprehensive_2022}, \cite{jiang_battery_2025}. Among available options, battery-based storage systems are considered among the most efficient forms of GSES due to their relatively low installation costs, high operational flexibility, and broad geographic applicability. Common challenges in developing battery technologies include mitigating safety hazards such as fire and explosion risks due to thermal runaway effects; reducing capital costs for installation; extending cycle life to maximize utilization and economic viability; ensuring responsible end-of-life disposal to prevent resource loss and environmental pollution; and guaranteeing system resilience across various temperatures, weather conditions, and disaster scenarios \cite{jiang_battery_2025}. On a broader scale, battery technologies must also address challenges of extreme conditions and compatibility with existing systems.

\paragraph{Resilience to Extreme Conditions}
Aerospace missions and extreme climate environments require energy systems to be able to operate under extreme environmental conditions, including extreme temperatures, which can severely degrade the performance and stability of electrochemical devices; ionizing radiation, which can damage storage systems, particularly those relying on capacitors or excited-state phenomena; and microgravity, which complicates phase separation in fluid-based systems and thus requires active separation mechanisms.

\paragraph{Compatibility}
In aerospace missions, common forms of energy generation include solar power, chemical fuels, and nuclear energy \cite{bushnell_frontiers_2021}. Compared to terrestrial systems, aerospace energy systems face additional constraints, such as strict payload limitations, which require energy systems to be compact and lightweight, while remaining robust enough to withstand high levels of vibration and acceleration during launch. If properly designed and implemented, regeneration systems could help optimize the size, weight, and cost of spacecraft energy systems. However, most current space systems lack energy regeneration capabilities \cite{bushnell_frontiers_2021}.

\subsubsection{Monitoring and Exploratory Technologies}
This category focuses on technologies designed for data collection, sensing, and navigation across both terrestrial and space environments. On Earth, these technologies are used in climate and environmental monitoring, helping address climate issues through comprehensive data tracking and collection. However, despite their popularity, technologies in this category face several challenges related to safety, security, policy regulation and liability, and compatibility. This section examines two key implementation challenges by focusing on two widely adopted technologies in this category: satellites and unmanned aerial systems (UAS).

\paragraph{Operability}
Satellite-based remote sensing has become an essential tool in Earth observation and environmental monitoring. Its applications range from climate surveillance to urban planning and disaster response \cite{Li2017-vj}\cite{Gong2024-ny}. Satellites face classic hardware challenges, including radiation exposure, thermal instability, limited payload capacity and power in miniaturized systems, and rising concerns over orbital congestion and space debris \cite{Chen2020-du}. As demand for higher-resolution and reliable spatial information grows, satellite systems must continue to evolve to meet both physical and computational demands.

\paragraph{Institutional Support}
The adoption of Unmanned Aerial Systems (UAS) has grown at an unprecedented rate in recent years, with widespread uptake across global markets \cite{Wargo2014-em}. These systems support applications such as aerial surveillance, emergency response, and communication services, particularly in hard-to-reach or hazardous environments \cite{Pal2024-xe}\cite{Telikani2025-hs}. Although UAS technologies face technical and operational challenges, they also encounter significant implementation barriers. The plethora of Federal Aviation Administration (FAA) rules outlining aircraft, pilot, maintenance standards, and flight restrictions makes the adoption of UAS operations costly, time-consuming, and bureaucratically difficult, thereby limiting their benefits and slowing their growth \cite{Wargo2014-em}.

\subsubsection{Novelty}
“Novelty” is defined as revolutionary innovations that fundamentally reshape how we approach complex challenges and act as catalysts, inspiring entirely new pathways, to play a critical role in space-sector R\&D. When a breakthrough proves viable in one high-demand context, it can often pioneer solutions in others, making it a strong BiT candidate. Mankins \cite{mankins_space_2009} recounts how NASA’s ambitious plans for annual robotic lunar missions were deeply cut back under the Exploration Systems Architecture Study, curtailing cutting-edge innovation in favor of lightweight, expendable architectures. Simultaneously, the NASA Institute for Advanced Concepts (NIAC), which served as a hub for visionary research, was disbanded. This marked a broader institutional pivot away from bold, high-risk exploration toward incremental, low-risk engineering. The result: significant delays, escalating mission costs, and a diminished capacity for long-term innovation. As Yung et al. \cite{yung_systematic_2021} observe, “the space sector … focuses mainly on advancing the technology conservatively rather than pursuing disruptive, radical breakthroughs,” stalling progress and narrowing future possibilities.

Despite the importance of novelty, its presence in the real world remains limited due to challenges from facing systemic barriers to large-scale deployment, to lack of scalability and sustainable business models, with lack of public awareness and acceptance and difficulty for policy compliance as major challenges. Without the ability to translate technical success into practical adoption, novelty remains stuck in the lab. Recognizing this, agencies have begun reinvesting in high-risk, high-reward programs like ESA’s Initial Support for Innovation (EISI), CNSA’s commercial space innovation consortium, and revived NIAC. They acknowledge that while most bold ideas may not succeed, those that do can redefine what is possible.
\paragraph{Public Awareness}
Radical technologies often struggle to gain momentum without demonstrated applications. In the absence of visible success or widespread awareness, public trust and investor confidence remain limited. Compounding this challenge, many early-stage technologies fall outside the traditional funding priorities of universities, private investors, or government agencies. Without stable financial and resource support, promising innovations frequently fail to transition from research to deployment. In addition, societal skepticism can further hinder adoption. Technologies that appear to challenge cultural norms (e.g. gene editing or lab-grown meat) may encounter resistance regardless of their technical merit. Ethical concerns, misinformation, or lack of transparency can delay implementation or spark public backlash.
\paragraph{Policy Compliance}
Novelty often outpaces regulation. Emerging technologies frequently operate in legal gray zones, where existing safety standards and liability frameworks are inadequate. A contemporary example is deep-sea mining, which has attracted significant investment, but its deployment is stalled by unresolved legal and environmental issues. Alam et al. \cite{alam_deep-sea_2025} argue that the risks significantly outweigh potential benefits if the activity operates under current regulations, highlighting the urgent need for governance mechanisms to catch up.

In addition to regulatory delay, inconsistent policy environments and international fragmentation pose further challenges, as well as administration change, budget cuts, and geopolitics. Moreover, governmental involvement introduces national safety and limited international collaboration, which is acceptable in defense contexts but counterproductive in areas like sustainability.

\subsection{Traits}
This section introduces eight traits identified as critical for the development of successful BiTs. Each trait corresponds directly to one of the eight challenges outlined in the preceding section, this translation highlights qualities that can be recognized early in the design process and cultivated throughout development. The traits and their working definitions are presented in Table \ref{tab:traits} below for readability. Together, they form the foundation of a set of recommendations intended to guide the creation and implementation of BiTs.

\begin{table}[ht]
\label{tab:traits}
\begin{center}       
\begin{tabular}{|l|p{12cm}|}
\hline
\rule[-1ex]{0pt}{3.5ex}  Trait & Description \\
\hline
\rule[-1ex]{0pt}{3.5ex} Circularity & The degree to which a technology minimizes waste across its lifecycle via reuse, repair, recycling, and material regeneration, thereby reducing environmental impact and reliance on finite inputs.\\
\hline
\rule[-1ex]{0pt}{3.5ex} Cost-Effectiveness & A technology’s ability to be produced, maintained, and scaled at a reasonable cost relative to its usage over the entire life cycle. \\
\hline
\rule[-1ex]{0pt}{3.5ex} Institutional Support & The extent to which governments, agencies, public or private institutions provide support for a technology through financial investment, policy advocacy, infrastructure support, or partnership facilitation. \\
\hline
\rule[-1ex]{0pt}{3.5ex} Operability & A technology’s ability to maintain performance with minimal human intervention, especially in remote, high-risk environments with reduced resources and limited access. \\
\hline
\rule[-1ex]{0pt}{3.5ex} Resilience & The ability of a technology to operate reliably under extreme environments, including but not limited to, high radiation, extreme temperatures, and pressure variations.\\
\hline
\rule[-1ex]{0pt}{3.5ex} Compatibility & A technology’s ability to integrate with current hardware and software systems via modular design or standard interfaces. \\
\hline
\rule[-1ex]{0pt}{3.5ex} Policy Compliance & The degree to which a technology aligns with applicable laws, regulations, and standards, thereby ensuring its responsible deployment. \\
\hline
\rule[-1ex]{0pt}{3.5ex} Public Acceptance & The degree to which stakeholders accept the usage of a technology, influenced by ethical, cultural, economic, social, and environmental concerns. \\
\hline
\end{tabular}
\vspace{10pt}
{\Large \caption{List of traits and their descriptions for Bidirectional Technologies.}}
\end{center}
\end{table}

\section{Actions and Recommendations}
\begin{figure*}[!h]
  \centering
  \includegraphics[width=\textwidth]{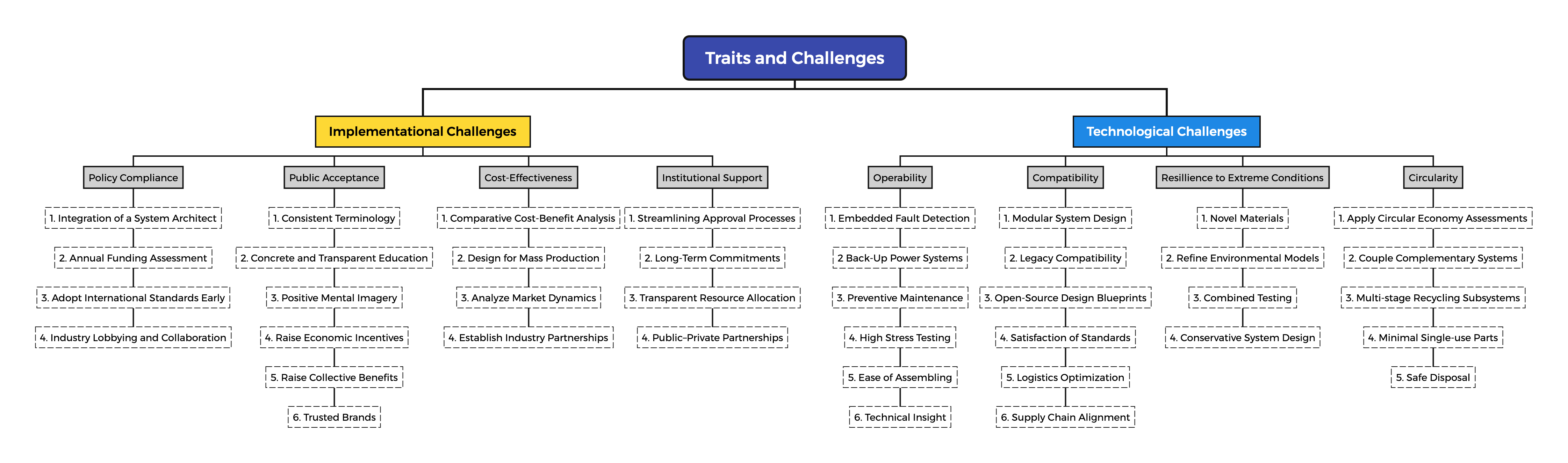}
  \caption{A summary of the recommendations.}
  \label{fig:recs}
\end{figure*}

In the previous section, eight key traits that address the challenges outlined in each category are identified. In this section, beyond recognizing desirable qualities, concrete solutions that can help achieve each trait are proposed. These actions are not meant to be prescriptive or exhaustive; instead, they serve as accessible starting points. This section is organized into eight subsections, each centered on one trait and its associated actions. At the end of the section, Figure \ref{fig:recs} is provided to summarize all the proposed actions.

\subsection{Circularity}
Drawing from Design for Repair-Reuse-Recyclability principles \cite{veerakamolmal_recycle_2000}, criteria in Life Cycle Assessment Tools \cite{di_noi_openlca_2017}, and existing circularity strategies \cite{esa_circular_2022, esa_closed-loop_2018}, this section provides actions to reduce the amount of waste released during the entire lifespan of a technology.
\begin{enumerate}
    \item \textit{Apply Circular Economy Assessments } - Apply circular economy assessment tools, such as life cycle analysis (LCA), material flow analysis (MFA), and input-output analysis (IOA), to continuously monitor waste.
    \item \textit{Couple Complementary Systems} -  Identify and couple complementary systems and processes where one system’s output can be reused as another’s input.
    \item \textit{Multi-stage Recycling Subsystems} – Incorporate hardware to capture, filter, and reuse waste outputs (water, air, etc.) so they become inputs again.
    \item \textit{Minimal Single-use Parts }- Using components that are refillable, rechargeable, or designed for long lifespans, instead of items that have to be discarded and replaced frequently.
    \item \textit{Safe Disposal} - For dangerous components, their disposal needs to be planned before production to prevent unnecessary contamination. 
\end{enumerate}

\subsection{Cost-Effectiveness}
Using principles from Design for Manufacturing \cite{odriscoll_design_2002} and market dynamics \cite{schelling_micromotives_2006}, this section offers actions to produce, maintain, and scale the technology at a reasonable cost.
\begin{enumerate}
    \item \textit{Comparative Cost-Benefit Analysis} - Compare the cost-benefit analysis of BiTs and Non-BiTs, enabling cost forecasting and helping to determine whether pursuing a BiT is worthwhile.
    \item \textit{Design for Mass Production} – Use standardized parts and modular sub-assemblies to simplify integration and document the processes for manufacturers to replicate.
    \item \textit{Analyze Market Dynamics} - Address differences between customer priorities, market trends, and product use cases to remain competitive in the market as a new product.
    \item \textit{Establish Industry Partnerships} - Collaborate with experienced manufacturers early on to secure reliable production, or set up production lines that can be expanded as needed.

\end{enumerate}

\subsection{Institutional Support}
The level and quality of institutional support, along with the efficiency of technology development, are closely linked \cite{clarke_how_2001} Drawing on papers that examine institutional support challenges, this section presents actions to strengthen BiTs implementation by applying assessment methods, providing roadmaps, enhancing transparency, and promoting collaboration opportunities \cite{clarke_how_2001, nelson_what_2008}.

\begin{enumerate}
    \item \textit{Streamlining Approval Processes} - Applying socio-technical system assessment methods such as Technology Innovation Systems (TIS) and the Multi-Level Perspective (MLP) can reduce bureaucratic delays by identifying and addressing bottlenecks in development processes and innovation systems.
    \item \textit{Long-Term Commitments} -  Incorporating technology development road maps, protects innovators from abrupt policy reversals and program instability. 
    \item \textit{Transparent Resource Allocation} -  Publishing funding rationales, and reviewer feedback provides predictable benchmarks, reduces corruption, and strengthens institutional legitimacy.
    \item \textit{Public–Private Partnerships} -  Building partnerships, consulting stakeholders, and involving universities and research centers increases legitimacy, shares knowledge and funding.

\end{enumerate}

\subsection{Operability}
Building on principles from \textit{Design for Manufacturability Handbook} in the section on Design for Serviceability, and from \textit{Engineering Design} in the section on Design for Maintainability \cite{bralla_design_1996, pahl_engineering_2007}, as well as the guidelines outlined in \textit{Reliability, Availability, and Serviceability} \cite{hsiao_reliability_1981}, this section proposes a series of actions to resolve challenges of reliably operate BiTs in harsh and remote climates: 
\begin{enumerate}
    \item \textit{Embedded Fault Detection} - A set of programmed sensors into the design that autonomously detects malfunctions and acts in critical, time-sensitive situations.
    \item \textit{Back-Up Power Systems} -  Built-in energy systems capable of sustaining critical operations and reporting subsystem status when the default power system is inoperative. 
    \item \textit{Preventive Maintenance}  - A set of tests that users can run to check the performance of individual components. 
    \item \textit{High Stress Testing} - Employ extreme testing programs (e.g., Design Fail Mode and Effects Analysis, Highly Accelerated Life Testing) that expose the technology to exceedingly extreme conditions, which reveal latent flaws that standard testing may overlook.
    \item \textit{Ease of Assembling} -  Design systems with mistake-proof techniques (e.g., keyed connectors, labelled parts) that require minimal technical skills to install. 
    \item \textit{Technical Insight} - Engage maintenance personnel and local users at multiple stages during the design phase to test mock-ups and identify risks related to spare parts, access, and repair tools. 

\end{enumerate}

\subsection{Resilience to Extreme Conditions}
This section relies on studies of design principles and challenges in harsh space environments \cite{balint_extreme_2008, nasa_radiation_1999}, and novel materials and system engineering designs \cite{eswarappa_materials_2023, qin_advancement_2021, yang_materials_2010, maurer_harsh_2008}. Four actions are proposed to improve the environmental resilience of BiTs, while also highlighting common examples where the strategy is highly relevant or urgent.

\begin{enumerate}
    \item \textit{Novel Materials} - Research and develop specialized materials with appropriate physical properties to withstand extreme conditions.
    \begin{enumerate}
        \item \textit{Novel Materials for High Temperature \& High Pressure Environments} - Optimize performance of alloy materials to improve their potential for industry applications.
        \item \textit{Novel Materials for Corrosive Chemicals} - Address issues brought forward by corrosion-susceptible materials and reduced corrosion-resistance due to recycling of materials. 
    \end{enumerate}
    \item \textit{Refine Environmental Models} - Build physically accurate models through data, sensitivity analysis, and uncertainty margins to evaluate hazardous environments. 
    \begin{enumerate}
        \item \textit{Risk Mapping for Collisions} - Use predictive models to identify locations of impacts and strengthen the corresponding structural integrity.
    \end{enumerate}
    \item \textit{Combined Testing} - Test systems under the combined, and preferably exaggerated, effects of multiple stressors that would exist concurrently. 
    \item \textit{Conservative System Design} - Design a system with redundancy for safety measures, built-in error correction, and safe shutdowns upon failure.
    \begin{enumerate}
        \item \textit{Multi-layered Design for Radiation} - Use a layered approach, including physical shielding and radiation-hardened materials, to address both cumulative ionizing effects and sudden disruptions.
    \end{enumerate}

\end{enumerate}

\subsection{Compatibility}
Drawing from principles from service-oriented architecture \cite{erickson_service_2009}, open-source hardware development \cite{bonvoisin_standardisation_2020}, and design-to-standards practices \cite{lee_design_1993}, this section proposes a series of actions to meet the implementational challenge of compatibility in BiTs: 
\begin{enumerate}
    \item \textit{Modular System Design} - construct components of BiTs as independent units, enabling swapping, reconfiguration, and adaptation of the technology without disrupting the overall system.
    \item \textit{Legacy Compatibility} -  Upgraded technology components remain compatible with previous legacy systems, ensuring that new modules can be integrated without requiring a full overhaul. 
    \item \textit{Open-Source Design Blueprints} - Publish modular schematics in open repositories to encourage collaboration across industry, academia, and citizen innovators.
    \item \textit{Satisfaction of Standards} - Designs should meet different widely accepted mechanical and electrical standards (e.g., ISO, ANSI, DIN) early in the design process to ease transitions between industries and nations.
    \item \textit{Logistics Optimization} - Design modules for ease of transport, to safeguards equipment with shock-absorbing designs to reduce transportation delays and costs.
    \item \textit{Supply Chain Alignment} - Have components easy to source and figure it out before mass production. Collaborate with suppliers to ensure they can meet demands

\end{enumerate}

\subsection{Policy Compliance}
Taking inspiration from the MIT System Architecture Group \cite{mit_defining_nodate} and the paper “Articulating the space exploration policy–technology feedback cycle” \cite{broniatowski_articulating_2008}, as well as the UNOOSA Space Modules \cite{unoosa_elearning_nodate}, this section proposes actions to the implementational challenge of policy compliance in BiTs.
\begin{enumerate}
    \item \textit{Integration of a System Architect} -  a coordinating actor capable of bridging the gaps between technical design and policy. 
    \item \textit{Annual Funding Assessment} - evaluate the technology program’s funding each year to ensure alignment with Executive and Legislative goals.
    \item \textit{Adopt International Standards Early} -  aligning early with established frameworks and space laws developing technologies reduces the risk of costly redesigns and helps secure approvals for export and deployment.
    \item \textit{Industry Lobbying and Collaboration} - Shared programs and open compliance tools can also lower costs and make the industry more predictable.
\end{enumerate}

\subsection{Public Acceptance}
Drawing from surveys and studies examining the public perception of novel technologies such as renewable energy sources \cite{gareiou_12_2021, sutterlin_public_2017}, and alternative meat \cite{vanloo_consumer_2020, chong_cross-country_2022, pakseresht_review_2022}, the following recommendations aim to promote the acceptance of BiTs to various stakeholders:
\begin{enumerate}
    \item \textit{Consistent Terminology} - Use consistent language across outreach and education efforts to establish familiarity and understanding of the technology.
    \item \textit{Concrete and Transparent Education} - Measure acceptance providing fully-disclosed information. Consult experts’ opinions and provide relevant credentials in educational efforts.
    \item \textit{Positive Mental Imagery} - Consult marketing experts and develop branding strategies to enhance the public appeal of the technology.
    \item \textit{Raise Economic Incentives} - Maximize visible economic benefits, such as creating employment opportunities, supporting local supply chains, and increasing local revenues.
    \item \textit{Raise Collective Benefits} - Improve and highlight collective benefits of the technology on the community, such as improved environmental impacts and energy security.
    \item \textit{Trusted Brands} - Companies may invest in partnerships with established brands and prioritize product launches in collectivistic societies to accelerate adoption and increase market share. 
\end{enumerate}

\section{Discussion}
\subsection{Limitations}
The recommendations outlined in Section 3 serve as an initial guide for designing and implementing BiTs, but they carry significant limitations. Given the novelty of BiTs as a field, the recommendations draw from a limited body of scholarly literature that lacks both empirical evidence and comprehensive datasets. As a result, it emphasizes traits and design principles rather than fixed benchmarks. At this stage, the recommendations  focus more on identifying patterns and possibilities than prescribing quantitative metrics of performance, efficiency, or cost.
A key omission is economic feasibility, which is a critical determinant of adoption and long-term success. Whether corporations and investors view BiTs as scalable and competitive depends heavily on market incentives such as subsidies, tax benefits, or sustainability mandates. These economic dynamics, while essential, all fall beyond the scope of this paper.
The recommendations focus on eight traits for early-stage design, which exclude other critical dimensions and traits. Governance structures, legal frameworks, intellectual property regimes, and ethical safeguards all shape how technologies are developed and deployed. Beyond these considerations, path dependency, the phenomena of suboptimal initial conditions constraining future outcomes, poses an additional barrier. Entrenched infrastructures and long-term investments in the space and climate sectors can lock in existing technologies and create friction for alternatives. These structural dynamics limit the flexibility needed for innovation, even when superior technologies are available. 
While this paper positions BiTs as catalysts for sustainable development, it would be naïve to ignore the potential for misuse. BiTs could exacerbate technological inequality if concentrated among wealthy nations or corporations. Such outcomes risk undermining the principles enshrined in the Space2030 Agenda. For these reasons, equity and regulations should be treated as a substantive design and policy priority, not an afterthought.
Finally, the recommendations are not intended to promote exploitation, violence, or militarization. Instead, it seeks to contribute to shared global goals of long-term sustainability and peace. BiTs should be understood as tools that can bridge domains, pool knowledge, and expand the range of solutions, provided they are guided by fairness, responsibility, and the peaceful use.

\subsection{Future Work}
With these problems and limitations in mind, there are abundant opportunities for future work to strengthen the implementation of BiTs.
A central priority is economic viability. Future studies should conduct systematic economic analyses that evaluate production costs, trade-offs against non-BiTs, and scaling pathways. Quantitative tools, such as cost–benefit analysis and life-cycle assessments, could provide concrete quantitative benchmarks. Recommendations should also be linked explicitly to funding mechanisms and public–private collaborations, with incentives adapted to different actors, from startups and small and medium sized companies (SMEs) to multinational corporations and government agencies.
The recommendations’ scope also calls for expansion. The eight proposed traits cannot account for universal BiT adoption. Moreover, the policy–technology interface, warrants closer study as national and regional space treaties, technology regulations, and political priorities could each accelerate or constrain BiT development. Equity research is equally urgent, identifying how BiTs can promote inclusive access and ensuring benefits extend beyond wealthy nations and corporations. At the same time, recommendations must be tailored to organizational scale. What works for NASA or the European Space Agency may not be feasible for SMEs or a developing country. Adapting guidance to these diverse contexts will enhance both the practicality and the global reach of BiTs. 
Finally, sustainability must be integrated into long-term planning. Circularity and end-of-life management require explicit attention to ensure BiTs do not shift environmental burdens elsewhere while claiming to address climate challenges. Future research should prioritize pathways for reuse, recycling, and safe disposal, particularly in settings where contamination presents pressing risks.

\section{Conclusions}
This paper presented a structured list of guidelines to facilitate the research and development of Bidirectional Technologies (BiTs), for the stakeholders. Through a systematic analysis of recurring challenges across four shared categories of terrestrial and space technologies, this paper found eight traits that are critical to the successful development and long-term adoption of BiTs. The guideline offers targeted strategies to address technical, institutional, and social barriers, while also highlighting areas for further research and investment.
Overall, the recommendations aim to serve as a reference for engineers, industry leaders, and policymakers to transition BiTs from emerging concepts into mature technologies. However, despite the promising potential of BiTs, their full integration into R\&D will require further advancements. Further analysis on technical, economic, and legislative aspects is essential. With these efforts, future work can expand the set of traits and strategies, offering a comprehensive framework to overcome barriers and maximize the impact.
This paper answers the call of global sustainability agendas, taking an initial step toward bridging Earth and space technologies. Step by step—or rather, BiT by BiT—these technologies can pave the way for a more sustainable tomorrow.

\section*{Acknowledgments}
The authors would like to thank the following individuals for their input and guidance in the development of this paper: Mr. Ryan Denmy, Configuration Lead for Crew Lander at Blue Origin, who mentored the initial stages of this work; Ms. Lily (Le Chuan) Wang, for her dedicated administrative leadership and her constructive, critical perspective that kept the authors’ research on track; and Mr. Arash Aslan-Beigi, a recent but highly valuable addition to the team, for his contributions to editing and drafting. The authors also extend their gratitude to the International Astronautical Federation for providing the opportunity and platform to present this research at IAC 2025 in Sydney, Australia.

No funding was received from the University of Toronto Aerospace Team (UTAT) or the University of Toronto to conduct this study.

\bibliographystyle{unsrt}
\bibliography{biblio}

\end{document}